\documentclass[letterpaper]{emulateapj}
\usepackage{apjfonts}

\usepackage{natbib}

\newcommand{\fig}[1]{\mbox{Fig.~\ref{#1}}}
\newcommand{\sect}[1]{\mbox{\S~\ref{#1}}}
\newcommand{\tab}[1]{\mbox{Table~\ref{#1}}}

\newcommand{\HST}{\mbox{\emph{HST}}}

\newcommand{\Lya}{\mbox{Ly$\alpha$}}
\newcommand{\Ha}{\mbox{H$\alpha$}}
\newcommand{\Hb}{\mbox{H$\beta$}}
\newcommand{\Hc}{\mbox{H$\gamma$}}
\newcommand{\Hg}{\mbox{H$\gamma$}}
\newcommand{\Hd}{\mbox{H$\delta$}}

\newcommand{\HII}{\mbox{H\,{\sc ii}}}
\newcommand{\MgII}{\mbox{Mg\,{\sc ii}}}
\newcommand{\fNII}{\mbox{[N\,{\sc ii}]}}
\newcommand{\fOII}{\mbox{[O\,{\sc ii}]}}
\newcommand{\fOIII}{\mbox{[O\,{\sc iii}]}}
\newcommand{\fSII}{\mbox{[S\,{\sc ii}]}}

\newcommand{\snr}{\mbox{S/N}}
\newcommand{\lam}{\mbox{$\lambda$}}

\newcommand{\LCDM}{\mbox{$\Lambda$CDM}}
\newcommand{\EBV}{\mbox{$E(\bv)$}}

\newcommand{\s}{\mbox{s}}
\newcommand{\Myr}{\mbox{Myr}}
\newcommand{\Gyr}{\mbox{Gyr}}

\newcommand{\ang}{\mbox{\AA}}
\newcommand{\cm}{\mbox{cm}}
\newcommand{\km}{\mbox{km}}
\newcommand{\kpc}{\mbox{kpc}}
\newcommand{\Mpc}{\mbox{Mpc}}

\newcommand{\gi}{\mbox{$g-i$}}

\defcitealias{mlf00}{MLF00}

\slugcomment{To appear in Astrophysical Journal}
\shorttitle{Infalling Faint \fOII\ Emitters in A851. I.}
\shortauthors{Sato \& Martin}

\begin{document}

\title{Infalling Faint \fOII\ Emitters in \object{Abell 851}.
I.~Spectroscopic Confirmation of Narrowband-Selected Objects}
\author{Taro Sato\altaffilmark{1} and Crystal
L. Martin\altaffilmark{2,3,4}} \affil{Department of Physics,
University of California, Santa Barbara, CA 93106-9530, U.S.A.}
\altaffiltext{1}{taro@physics.ucsb.edu}
\altaffiltext{2}{cmartin@physics.ucsb.edu}
\altaffiltext{3}{Packard Fellow}
\altaffiltext{4}{Alfred P. Sloan Foundation Fellow}

\begin{abstract}
We report on a spectroscopic confirmation of narrowband-selected
\fOII\ emitters in \object{Abell 851} catalogued by \citet{mlf00}.
The optical spectra obtained from the Keck~I Low Resolution Imaging
Spectrometer (LRIS) and Keck~II Deep Imaging Multi-Object Spectrograph
(DEIMOS) have confirmed $\fOII\lam3727$ emission in
narrowband-selected cluster \fOII\ candidates at a $\approx 85\%$
success rate for faint ($i \la 25$) blue ($\gi < 1$) galaxies.  The
rate for the successful detection of \fOII\ emission is a strong
function of galaxy color, generally proving the efficacy of narrowband
\fOII\ search supplemented with broadband colors in selecting faint
cluster galaxies with recent star formation.  Balmer decrement-derived
reddening measurements show a high degree of reddening [$\EBV \ga
0.5$] in a significant fraction of this population.  Even after
correcting for dust extinction, the $\fOII/\Ha$ line flux ratio for
the high-\EBV\ galaxies remains generally lower by a factor of $\sim
2$ than the mean $\fOII/\Ha$ ratios reported by the studies of nearby
galaxies.  The strength of \fOII\ equivalent width shows a negative
trend with galaxy luminosity while the \Ha\ equivalent width does not
appear to depend as strongly on luminosity.  This in part is due to
the high amount of reddening observed in luminous galaxies.
Furthermore, emission line ratio diagnostics show that AGN-like
galaxies are abundant in the high luminosity end of the cluster
\fOII-emitting sample, with only moderately strong \fOII\ equivalent
widths, consistent with a scenario of galaxy evolution connecting AGNs
and suppression of star-forming activity in massive galaxies.
\end{abstract}



\section{Introduction}

The currently popular cosmological paradigm predicts that galaxy
clusters, which are the largest gravitationally bound systems, evolve
by merging and accreting smaller structures from the surrounding field
region \citep{kau95a,kau95b}.  Results from recent simulations have
shown that galaxy clusters in \LCDM\ cosmology have two rather
distinct phases in their mass assembly history, merger-dominated and
accretion-dominated phases \citep{zha03}.  While the star-forming activities in the universe are generally declining since
$z \sim 2$ \citep{lil96,mad96}, some galaxy clusters may remain
dynamically active at intermediate redshifts of $z \sim 0.5$
\citep{tas04}, potentially triggering cluster-scale enhancement of
star-forming activity.  The dynamic range in local
environments that galaxy clusters offer has been an essential aid for
the study of galaxy evolution, particularly of star-forming field
spirals into passively-evolving cluster spheroidal galaxies, spanning
a wide range of morphological types in the Hubble sequence.

As the connection between structure formation and galaxy clusters
becomes clearer, most recent efforts in cluster galaxy surveys have
focused on wide-field imaging in an attempt to trace the substructures
extending from their central regions out to larger cluster radii,
where galaxies are not yet virialized within strong cluster
gravitational potentials \citep{abr96b,kod05}.  Such recent studies
have shown that the mechanism responsible for the transformation of
cluster galaxies must be in effect well beyond the cluster virial
radius, well before the galaxies at infall reach the central region of
a cluster \citep[e.g.,][]{bal97,tre03,mor05}.

\citet[][hereafter \citetalias{mlf00}]{mlf00} used narrowband and
broadband photometry to select faint $\fOII\lam3727$ emission-line
candidates in a rich, moderately distant ($z \simeq 0.4069$) galaxy
cluster \object{Abell 851} (Cl0939+4713; $9^{\rm h}42^{\rm m}56\fs2$,
$46^\circ59'12\farcs0$).  The cluster offers an especially interesting
laboratory for the study of cluster galaxy evolution, because there is
observational evidence of large-scale filaments detected via
photometric redshift technique by \citet{kod01}, giving observers a
snapshot of the on-going assembly of the cluster.  Even before the
identification of the filaments, \object{Abell 851} had drawn
attention due to its high fraction of blue and post-starburst galaxy
population \citep{dre99} compared to other clusters at similar
redshifts \citep{lot03}.  These observations are apparently related to
the X-ray observations indicating highly dynamically active state of
\object{Abell 851} \citep{schi96,schi98,def03}.

\citetalias{mlf00} narrowband search yielded $371$ cluster
emission-line member candidates over the projected area of
$\sim~181~{\rm arcmin}^2 \approx 4.4~\rm{Mpc}\times4.4~\rm{Mpc}$ at
the cluster redshift.  The completeness limit of the narrowband survey
was $m_{5129}({\rm AB}) \simeq 24$ for galaxies with \fOII\ emission
equivalent width of $\la -11~\ang$, making it among the faintest
surveys of the kind at $z \sim 0.4$.  The locus in the $(\gi)$-$i$
color-magnitude diagram and their small isophotal area suggested that
a significant number of the \fOII\ emission candidates were dwarf
galaxies.  Dwarf and large spiral galaxies are expected to respond
differently to cluster tidal fields due to their different mass
concentrations \citep{moo99}.  The efforts spanning a few decades have
established such empirical relations as the morphology-density
relation \citep{dre80} and the Butcher-Oemler effect \citep{but84} for
the bright, giant cluster galaxies, but the underlying physical
processes that are responsible for the evolution of galaxy properties
and therefore the aforementioned empirical relations have yet to be
well constrained.  The sign of the differential evolution between
dwarf and giant galaxies is even more obscure, since few wide-field
surveys to date have sampled faint galaxies at high redshifts.

As a spectroscopic follow-up of \citetalias{mlf00}, we have obtained
an extensive set of spectra of the \fOII-emitting candidates using the
Keck~I Low Resolution Imaging Spectrometer \citep[LRIS;][]{oke95} and
Keck~II Deep Imaging Multi-Object Spectrograph
\citep[DEIMOS;][]{fab03}.  In a series of papers drawing on this
spectroscopic sample, we wish to study infalling faint star-forming
galaxies in a highly dynamically active galaxy cluster at intermediate
redshift.  We present the observation, data reduction, and various
measurement methods employed in this study in \sect{sec.Data}.

Although galaxy clusters are well localized within redshift space and
much denser in their projected galaxy number density compared to that
of the surrounding field, an inherent observational challenge in
wide-field study of clusters is to obtain secure membership
identification of cluster galaxies.  Various photometric methods exist
for this purpose, each sensitive to different spectral features of
distant objects.  Narrowband photometry is a useful method of
selecting emission-line objects within a wide field of view over a
redshift range defined by the filter bandpass.  While efficient for
cluster galaxy surveys, narrowband selected objects always need
further constraints on their true identity via other means, since the
selection method is prone to detecting emission lines or prominent
spectral breaks from foreground or background objects at different
redshifts.  Therefore in \sect{sec.ConfirmationOfOIIEmitters} we
assess the efficacy of narrowband technique in our narrowband-selected
sample of galaxies.  In the same section we also explore possible
systematic errors introduced by using two independent sets of
measurements, narrowband photometry and spectroscopy.

Studies of intermediate redshift galaxies often use $\fOII\lam3727$
emission line as a tracer of star-forming activity for its
availability within the optical wavelengths out to $z \sim 1$.  The
justification for the use of the \fOII\ emission line as a star
formation tracer originates from the apparent empirical correlation
found between $\fOII\lam3727$ flux and \Ha\ recombination line flux
within the sample of galaxies with various morphological types in the
local universe \citep{gal89,ken92}.  Low-order Balmer recombination
lines are presumed to be good star formation indicators because of
their tight coupling to the UV radiation intensity from young O- and
B-type stars in \HII\ regions.  The correlation between the \fOII\ and
the Balmer recombination lines, however, is complicated by its
dependence on the elemental abundances and ionization state of the
parent \HII\ region.  The $\fOII\lam3727$, being on the bluest side of
optical wavelengths, also suffers greater dust extinction.  As dust
reprocessing of starlight appears to be more important in high
redshift galaxies \citep[e.g.,][]{hic02,lef05}, the reliability of the
\fOII\ emission as a star formation tracer, especially for high
redshift galaxies, still remains an open topic \citep{jan01}.  For
this reason, we study in \sect{sec.PropertiesOfTheClusterOIIEmitters}
the properties of \fOII\ emitters as observed in our sample.  We then
summarize our findings in \sect{sec.Summary}.

Throughout this paper, we use the standard \LCDM\ cosmology,
$(\Omega_m, \Omega_\Lambda) = (0.3, 0.7)$, with $H_0 =
70~\km~\s^{-1}~\Mpc^{-1}$.  At the distance of \object{Abell 851} ($z
= 0.4069$), $1''$ on the sky corresponds to a projected physical
distance of $5.43~\kpc$, and the lookback time to the galaxy cluster
is $4.3~\Gyr$.

\section{Data}
\label{sec.Data}

\subsection{Observations \& Data Reductions}
\label{sec.ObservationsAndDataReductions}

Spectra of the emission candidate members of \object{Abell 851} were
obtained using the Keck~I LRIS and Keck~II DEIMOS on the nights of
2002 January 6-8 and 2003 January 26-27, respectively.  These
observing runs were made in photometric conditions, with seeing of
$0\farcs7$--$0\farcs8$, measured with two-dimensional spectra of
standard stars; the seeing appeared slightly worse during DEIMOS
observations.

All the LRIS spectra analyzed in this paper were obtained using the
D460 dichroic with the 600/4000 grating.  The spectral coverage was
$\sim 5000 - 7000~\ang$, enough for the inclusion of the redshifted
$\fOII\lam3727$ up to $\fOIII\lam5007$, but not \Ha\ for the objects
at the redshift of interest.  There were six different multislit
masks, each with $20 - 25$ slits.  The slit width was fixed at $1''$,
yielding the spectral resolution of $\simeq 4.8~\ang$~FWHM of arc
lines.  For most masks, the total exposure time was $7200$ seconds.
Our LRIS data were reduced using the standard IRAF\footnote{IRAF
(Image Reduction and Analysis Facility) is distributed by the National
Optical Astronomy Observatories, which are operated by AURA, Inc.,
under cooperative agreement with the National Science Foundation.}
facilities along with custom PyRAF\footnote{PyRAF is a product of the
Space Telescope Science Institute, which is operated by AURA for NASA.
} scripts for the most common tasks.  The individual frames were bias
subtracted and corrected for pixel-to-pixel sensitivity variations.
The science frames for each mask were then coadded with the cosmic-ray
rejection option after being registered using the positional
cross-correlation of bright emission features.

All DEIMOS spectra were obtained using the $600$ lines/mm grating and
the GG495 filter.  A total of five multislit masks, each with $40-50$
$1''$-slits, were used.  The spectral resolutions were $\simeq
2.7~\ang$ and $\simeq 3.2~\ang$~FWHM of arc lines for blue and red
sides, respectively.  The total exposure times were $9000$, $9000$,
$10800$, $5195$, and $10800$ seconds for these masks.  The DEEP2
DEIMOS data pipeline\footnote{The analysis pipeline used to reduce the
DEIMOS data was developed at UC Berkeley with support from NSF grant
AST-0071048.}  was used to produce final one-dimensional spectra of
DEIMOS observations.  Before flux calibration, these one-dimensional
spectra were converted with a purpose-written Python program into FITS
images that can be processed on IRAF or PyRAF, with which most
subsequent data analysis were made.  All one-dimensional spectra were
linearized and rebinned to $1.28~\ang$ and $0.65~\ang$ per pixel for
the LRIS and DEIMOS spectra, respectively.

We also make use of Sloan $g$ and $i$ photometry from the
\citetalias{mlf00} imaging survey.\footnote{ There was a $+0.43$ mag
error in the \citetalias{mlf00} $g$ photometry as a result of an error
in the previously published \object{Abell 851} $g$ photometry
\citep{dre92} used for their original calibration \citep{lot03}.  In
this paper, we have recalibrated our $g$ photometry as in
\citet{lot03}, such that $\gi = (\gi)_{\rm MLF00} - 0.43$. }  For the
detail on the broadband photometry, see \citetalias{mlf00}.

All target spectra were flux-calibrated with Kitt Peak
spectrophotometric standards~\citep{mas88}.  In the LRIS observations,
the standards were observed with a longslit mask ($1.5'' \times
175''$).  For the DEIMOS run, a slitless spectrum was obtained for
each standard star.  We did not correct for foreground Galactic
reddening toward the target field, as it is marginal at $\EBV \simeq
0.016$ \citep{sch98} compared to the average intrinsic reddening of
target galaxies.

In reality, absolute flux calibration does not necessarily yield a physically reliable estimate of source flux from the region of
interest due to various observational systematics.  The spatial
coverage of a target depends on the aperture size relative to the
source, causing aperture effects.  A slight offset in the alignment of
slit mask during different exposures introduces systematics as well.
The alignment effect appeared relatively small, as the repeat
observations with the same instrument yielded the variation within a
factor of $\simeq 2$ in the mean continuum flux densities within the
rest-frame $4150-4300~\ang$ (\sect{sec.SpectralLineMeasurements}).
Among the repeat observations with the different instruments, the mean
continuum flux was smaller by a factor of $\simeq 2$ for a vast
majority, but not all, of DEIMOS spectra.  Since both the LRIS and
DEIMOS observations were made with the same aperture size in
photometric conditions, we suspect the slit alignment may have been
worse in the DEIMOS observations; slightly worse seeing during the
DEIMOS run may have also contributed to the discrepancy.

\begin{deluxetable}{lccc}
\tablecaption{Spectroscopic Target Selection}
\tablehead{
\colhead{} &
\colhead{Spectroscopic} &
\colhead{\fOII\ Emission} &
\colhead{Correction} \\
\colhead{Color} &
\colhead{Targets} &
\colhead{Candidates} &
\colhead{Factor}
}
\startdata
$\gi < 1$ & $71$ & $100$ & $1.41$ \\
$1 < \gi < 2$ & $89$ & $135$ & $1.52$ \\
$2 < \gi$ & $31$ & $81$  & $2.61$ \\
N/A\tablenotemark{a}  & $18$ & $55$  & $3.06$
\enddata

\tablecomments{ Correction factors are the number of \fOII\ emission
candidates divided by the number of spectroscopic targets. }

\tablenotetext{a}{ Objects with their uncertainty in \gi\ color
exceeding $0.5$. }
\label{tab.TargetSelection}
\end{deluxetable}

In \citetalias{mlf00} the \fOII\ emission candidates were selected
based on the \snr\ of on-band flux excess through the on-band
narrowband filter, with the continuum narrowband filter blueward of
on-band.  The choice of this specific narrowband filter was made based
on the redshift ($z = 0.4067$) and cluster velocity dispersion
($\sigma_v = 860~\km~\s^{-1}$) reported by~\citet{dre92} to ensure
that the redshifted $\fOII\lam3727$ emission line from a majority of
the virialized cluster members fall within its passband.  A total of
$371$ objects with the on-band flux excess at $\snr > 3$ level made
the category of \emph{on-band-selected sample} of emission-line galaxy
candidates.  Due to the limited wavelength coverage of LRIS data, our
DEIMOS run had a significant number of repeat observations of LRIS
objects to ensure simultaneous detection of \fOII\ and \Ha\ emissions.
A total of $212$ on-band-selected targets were observed in this
spectroscopic follow-up; in three of these targets we could not detect
significant continuum, reducing the sample size to $209$.

The spectroscopic targets, drawn from the \citetalias{mlf00}
on-band--selected sample, were mostly selected by their \gi\ color,
favoring bluer objects; no further \snr\ cut was made within this
spectroscopic sample.  \citetalias{mlf00} divided the \fOII\ emission
candidates into three \gi\ color categories: blue ($\gi < 1$), green
($1<\gi<2$), and red ($\gi>2$).  The objects with poor photometry
(i.e., the uncertainty in \gi\ color exceeding $0.5$) are excluded
from these categories, constituting ``gray'' (in \citetalias{mlf00})
or N/A objects.  Our initial intention was to make the $\gi<1$ sample
as complete as possible, since true \fOII\ emitters are expected to
have a blue \gi\ color in general.  However, a calibration error in
the original photometry shifted many objects into the $\gi<1$ category
for this spectroscopic follow-up, making our $\gi < 1$ sample larger
in number but less complete (\tab{tab.TargetSelection}).

If repeat observations existed, the spectrum with the highest
continuum \snr\ (\sect{sec.SpectralLineMeasurements}) was generally
used for analysis (i.e., the \emph{highest-\snr\ spectrum}) unless a
significant defect was noted for that spectrum upon visual inspection,
in which case the spectrum of the second highest \snr\ was used.  At
times we supplement our data by using a lower-\snr\ spectrum if the
spectral range of the highest-\snr\ spectrum does not cover spectral
features of interest; we indicate the use of such \emph{secondary
spectrum} where appropriate.

\subsection{Spectral Line Measurements}
\label{sec.SpectralLineMeasurements}

\begin{figure}
\plotone{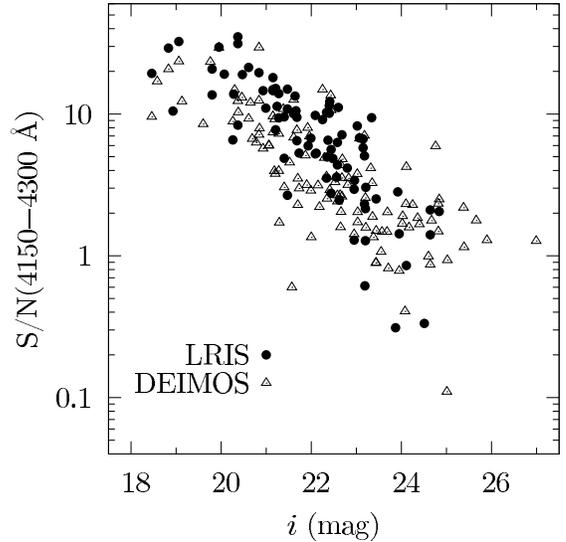}
\caption{ Continuum \snr\ per pixel of spectra as a function of their
source $i$ magnitude.  See \sect{sec.SpectralLineMeasurements} for the
definition of continuum \snr.  }
\label{fig.ContSNRVsIMag}
\end{figure}

To compute characteristic signal-to-noise ratio (\snr) of a spectrum,
a 3rd-order polynomial was fitted to the continuum within the
rest-frame wavelength range of $4150 - 4300~\ang$ with two iterations
of sigma clipping at $\pm 2\sigma$, where $\sigma$ is the RMS residual
noise per pixel.  The mean flux density of the fit within the spectral
range was divided by $\sigma$ to obtain the continuum \snr\ per pixel.
The resulting continuum \snr\ distribution is shown in
\fig{fig.ContSNRVsIMag}.  A large number of spectra of varying
qualities prompted us to standardize the measurement process with
purpose-written software employing two methods of line measurement,
suitable for different purposes.

\subsubsection{Gaussian-Fitting Method}
\label{sec.GaussianFittingMethod}

\begin{deluxetable}{cccc}
\tablecaption{Spectral Line Index Definitions}
\tablewidth{0pt}
\footnotesize
\tablehead{
\colhead{Line} &
\colhead{Blue continuum} &
\colhead{Line} &
\colhead{Red continuum}\nl
\colhead{} &
\colhead{(\AA)} &
\colhead{(\AA)} &
\colhead{(\AA)}
}
\startdata
\fOII\lam3727 & 3653--3713 & 3713--3741 & 3741--3801\\
\Hd & 4030--4082 & 4082--4122 & 4122--4170\\
\Hg & 4230--4270 & 4320--4359 & 4367--4383\\
\Hb & 4785--4815 & 4821--4901 & 4925--4949\\
\fOIII\lam5007 & 4960--4985 & 4993--5021 & 5030--5055\\
\Ha & 6493--6532 & 6553--6574 & 6596--6636\\
\fNII\lam6583 & 6493--6532 & 6574--6593 & 6596--6636
\enddata
\label{tab.SpectralLineIndexDefinitions}
\end{deluxetable}

For each spectral line, a straight line was fitted to the continuum
flux blueward and redward of a line to estimate the continuum flux
density.  The linear baseline was then subtracted from the spectral
line region, over which either one or two Gaussian profiles were
fitted to the spectrum, depending on the presence of emission,
absorption or both lines.  These spectral ranges are predefined
(\tab{tab.SpectralLineIndexDefinitions}).  The results of the
fit were inspected visually for all the spectral features, and the
spectral ranges were often adjusted to avoid data points obviously
affected by anomalies such as telluric features and cosmic rays.

Some Balmer lines, e.g., \Hb, \Hc, and \Hd, often show both emission
and absorption lines.  In such cases, emission and absorption line
fluxes were computed separately from the best fit parameters of each
Gaussian.  The uncertainty in the line flux thus obtained was
estimated from the RMS of the Gaussian profile over the spectral line
region, multiplied by $\sqrt{N \Delta\lambda}$, where $N$ is the
number of data points used in the fit, and $\Delta\lambda$ is the
dispersion of the spectrum per pixel.

There is an intrinsic difficulty in separating the nebular and stellar
components in Balmer lines.  As described above, these components were
fitted with two Gaussians when they were visually identified.  For
low-\snr\ spectra, this was often not possible, and only an emission
or absorption feature was fitted.  The Balmer absorptions in the
stellar atmosphere are presumably always present, however, and we
underestimate an emission flux in presence of undetected absorption
flux (or vice versa).  When appropriate, we corrected emission flux
artificially by adding an equivalent width of $2~\ang$ at \Ha, \Hb,
and \Hc\ \citep{mcc85}, if absorption features were not explicitly
fitted.

\subsubsection{Flux-Summing Method}
\label{sec.FluxSummingMethod}

The continuum baseline near a line was estimated as in the
Gaussian-fitting method (\sect{sec.GaussianFittingMethod}).  The line
strength was measured by summing the differences between the observed
spectrum and the continuum baseline over the spectral line region.
The uncertainty in this strength was estimated by summing in
quadrature the values stored in the sigma vector of 1D spectrum and
the RMS of continuum fit for all pixels in the spectral line region.
The measurements through this method were done uninteractively using
the redshifts computed from the line centers measured from the
Gaussian-fitting method (\sect{sec.RedshiftDetermination}).  The uncertainty in the continuum flux at the line center was determined
from the RMS of linear fit divided by the square root of number of
data points used in the continuum fitting.

For both the above methods, the line or continuum fitting was
done by a standard nonlinear least-square regression routine with
variance weighting using the formal sigma vector computed for each 1D
spectrum in the data reduction process.  For each line feature, its
equivalent width $W$ was computed from
\[
W = - \frac{F}{F_\lambda} \ ,
\]
where $F$ is the line flux and $F_\lambda$ is the continuum flux
density at the line center.  We use the sign convention where $F > 0$
for lines in emission and $F < 0$ for lines in absorption; therefore
the equivalent width is negative for a pure emission line and is
positive for a pure absorption line.  We use this sign convention
throughout the paper.  At times sticking to this convention can cause
confusion, since we often use a word increase (decrease) to mean the
\emph{magnitude} of equivalent width increases (decrease), deviating
from its mathematical sense.  For example, an equivalent width growing
in emission means the value is getting more negative.  Therefore we
caution that contextual interpretation is sometime necessary when the
behavior of equivalent widths is discussed, i.e., we may often not use
expressions that strictly follow mathematical senses.  We distinguish
the observed and rest-frame quantities by the use of subscript with
zero, e.g., $W$ and $W_0$ are observed and rest-frame equivalent
widths, respectively.

Note that the equivalent width from the flux-summing method estimates
the total flux in a line, i.e., the sum of emission \emph{and}
absorption fluxes, relative to the continuum flux density at the line
center.  This method has a disadvantage that the distinction between
nebular emission and stellar absorption, for example, cannot be made
from its measurements even when these features are well resolved.
Although the spectral dispersion was good enough to resolve emission
and absorption lines in many cases, there were difficulties
identifying spectral features for low-\snr\ spectra.  The
Gaussian-fitting method, requiring visual attention, is not effective
for those spectra for obvious reasons.  In addition, the effect of
emission filling in absorption features may not be apparent in
low-quality spectra.

\begin{figure}
\plotone{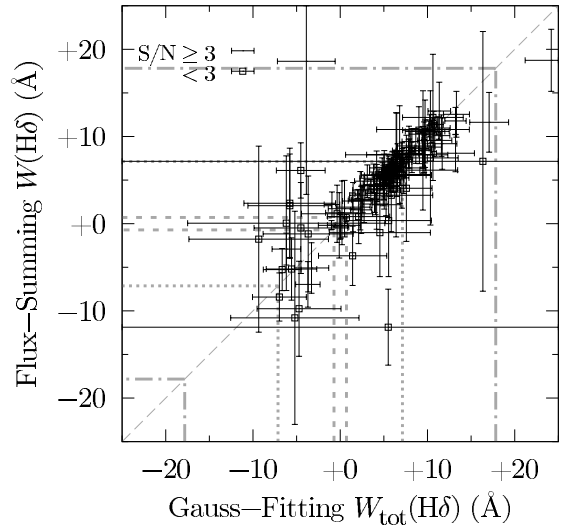}
\caption{ \Hd\ equivalent widths from two different measurement
methods (\sect{sec.SpectralLineMeasurements}).  $W_{\rm tot}(\Hd)$ is
computed from the sum of emission and absorption equivalent widths
using the Gaussian-fitting method.  Open squares denote measurements
for which $\snr < 3$, where \snr\ here is the flux density at line
center divided by the RMS noise derived from the straddling continua,
i.e., a measure of spectral quality insensitive to the strength of
\Hd\ line flux.  The dashed gray line indicates loci of equality.  The
dashed, dotted, and dash-dotted lines indicate the minimum (absolute)
equivalent widths that can be measured with \snr\ of $10$, $1$, and
$0.4$, respectively, at the LRIS dispersion, $1.28~\ang$ per pixel.  }
\label{fig.MgaussVsMfsumWHd}
\end{figure}

The flux-summing method does not directly yield physically meaningful
measurements when line features are complex, but has an advantage in
that faint features can be measured statistically.  Hence the
flux-summing method is suitable for indexing spectral features, being
less dependent on the quality of spectra.  The Gaussian-fitting method
is the preferred method for extracting physically meaningful
information such as line ratios used in emission-line diagnostics.
The two methods generally yield consistent results except for very
low-\snr\ measurements (\fig{fig.MgaussVsMfsumWHd}).

Hence we used the flux measurements from the Gaussian-fitting method
if both emission and absorption are present at a line and need to be
separated to each component, which is the case when emission-line flux
ratios involving Balmer lines are desired, for example.  Since our
analysis mostly is on emission line properties, the Gaussian-fitting
method extracts the emission component when absorption is also
present.  With low-\snr\ spectra or very emission-dominated line (such
as \Ha) for which visually fitting the two components are not
possible, we assumed a fiducial amount of underlying absorption flux
to be present, and corrected emission flux accordingly
(\sect{sec.GaussianFittingMethod}).  We used the flux-summing method
mostly to \emph{index} spectral properties, e.g., placing galaxies in
\fOII-\Hd\ equivalent width plane to identify post-starburst galaxies
in our forth-coming paper.  In other words, we did not use flux-summed
measurements to push measurements toward very faint emission line flux
in general.\footnote{An important exception is made for a pure
emission line such as $\fOII\lam3727$, for which no absorption
correction is necessary.}

\subsection{Redshift Determination}
\label{sec.RedshiftDetermination}

The redshift was estimated from the central wavelength of all the line
profiles using the Gaussian-fitting method.  Since emission and
absorption lines were fitted separately when both were visually
identified, the redshifts were computed separately for them as well.
The best estimate and its standard deviation are simply computed from
the list of redshifts thus obtained for each spectrum.  Although no
weights were assigned for the line features, any redshift measurement
beyond $1\sigma$ were rejected before the final determination of
redshift for that spectrum.

\subsubsection{Cluster Membership}
\label{sec.ClusterMembership}

In \citetalias{mlf00} narrowband survey, cluster membership was simply
assumed for the objects for which a candidate $\fOII\lam3727$ emission
line was detected within the on-band filter.  Note that more
up-to-date estimates of cluster velocity dispersion give higher
$\sigma_v$ for \object{Abell 851} than the one used to design the
original survey; at $\sigma_v = 1260~{\rm km~s}^{-1}$ \citep{dre99},
about $75\%$ of virialized cluster members would have been detected,
which is a lower fraction than originally intended.  We classified an
object to be a cluster member if its redshift was within $0.395 < z <
0.420$, corresponding to the on-band filter used in \citetalias{mlf00}
\fOII\ narrowband search (approximately $5198.7 - 5290.6~\ang$).

\section{Confirmation of \fOII\ Emitters}
\label{sec.ConfirmationOfOIIEmitters}

\citetalias{mlf00} discussed the likelihood of falsely detecting an
emission line or steep continuum profile of foreground or background
object in their narrowband search.  Their conclusions were as follows:
(1)~While the $4000~\ang$ break of the foreground early-type galaxies
would be a major cause of false detection in the \fOII\ narrowband
search, they could be distinguished by their red \gi\ colors; (2)~the
small survey volume meant prominent $\fOIII\lam\lam4959,5007$ doublet
and the \Hb\ emission of low-$z$ interlopers would produce only a
small number of false detection; and (3)~considering the competing
effects of increasing survey volume and decreasing apparent
brightness at high $z$, background \Lya\ emitters were not likely to
produce a significant number of false detections.

\subsection{True Detection Rate}
\label{sec.TrueDetectionRate}

To evaluate the efficacy of \fOII\ narrowband search, we used a subset
of spectra for which the on-band filter passband ($5198.7 -
5290.6~\ang$ or $0.395 < z < 0.420$ for $\fOII\lam3727$) was fully
within the observed spectral range for that spectrum, i.e., the
aperture location on a slit mask and/or dispersion did not move the
\citetalias{mlf00} on-band out of usable portion of CCD chip.  Based
on this criterion, we rejected $29$ targets from the $209$
spectroscopic targets, reducing the total number of galaxies to $180$.

We used a simple criterion for a true detection of \fOII\ emission.
For an object to be classified as a confirmed \fOII\ emitter, it had
to have $W(\fOII) < -11~\ang$ measured both in the Gaussian-fitting
and in the flux-summing methods (\sect{sec.SpectralLineMeasurements});
the equivalent width cut roughly corresponds to the narrowband survey
detection limit of \citetalias{mlf00}.  The spectrum that did not
satisfy this criterion either was of a very low-\snr\ or would
indicate a true non-detection in our definition; the distinction was
made such that the true non-detection of \fOII\ for an object meant
that its spectroscopic \snr\ was high enough that $W(\fOII) <
-11~\ang$ could have been detected at $2\sigma$ level.  To
better reproduce the observing condition of \citetalias{mlf00} survey,
the continuum \snr\ was computed using the RMS noise within the
wavelengths corresponding to the \citetalias{mlf00} on-band filter
passband after three iterations of $\pm2\sigma$ clipping.

\begin{deluxetable*}{lccccccccc}
\tabletypesize{\footnotesize}
\tablewidth{0pt}
\tablecaption{\fOII\ True Detection Rate by \gi\ Color }
\tablehead{
\colhead{} &
\colhead{} &
\multicolumn{2}{c}{No Redshift} &
\colhead{} &
\multicolumn{4}{c}{With Redshift} &
\colhead{} \\
\cline{3-4}
\cline{6-9} \\
\colhead{Color} &
\colhead{$N$\tablenotemark{a}} &
\colhead{Low \snr\tablenotemark{b}} &
\colhead{Other\tablenotemark{c}} &
\colhead{} &
\colhead{Low \snr\tablenotemark{b}} &
\colhead{Non-Member\tablenotemark{d}} &
\colhead{No \fOII\tablenotemark{e}} &
\colhead{True \fOII\tablenotemark{f}} &
\colhead{Rate (\%)}
}

\startdata
$\gi < 1$            & 63 &  9 &  3 & & 0 & 5 & 0 & 46 & 85.2 \\
$1 < \gi < 2$        & 80 &  5 & 11 & & 0 & 7 & 7 & 50 & 66.7 \\
$2 < \gi$            & 24 &  5 & 16 & & 0 & 0 & 0 &  3 & 15.8 \\
N/A\tablenotemark{g} & 13 &  6 &  2 & & 0 & 0 & 0 &  5 & 71.4
\enddata

~\tablenotetext{a}{Total number of \citetalias{mlf00} \fOII\ emission
candidates included in the true detection rate analysis. }

~\tablenotetext{b}{On-band spectroscopic \snr\ too low to detect
\fOII. }

~\tablenotetext{c}{High enough on-band \snr\ for \fOII\ detection but
the line visually undetected. }

~\tablenotetext{d}{Measured redshift moved \fOII\ outside the on-band
filter passband.}

~\tablenotetext{e}{\fOII\ expected within the on-band from the
measured redshift but not detected.}

~\tablenotetext{f}{$W(\fOII) < -11~\ang$ detected within the
on-band. }

~\tablenotetext{g}{ Objects with their uncertainty in \gi\ color
exceeding $0.5$.}

~\tablecomments{ True detection of \fOII\ emission required $W(\fOII)
< -11~\ang$ at $2\sigma$ level.  Low-\snr\ targets were excluded in
the computation of true detection rate (last column).  }

\label{tab.OIITrueDetectionsByGIClr}
\end{deluxetable*}

In \tab{tab.OIITrueDetectionsByGIClr}, we show how the true detection
rates break down by \gi\ color.  The subset of sample used for the
analysis is first divided into those with redshift measurements and
without.  The subset with no redshifts consists of two types of
objects.  Since our primary interests were \fOII\ emitters, we did not
measure the line features of apparently early-type spectra, leaving
redshifts unknown for most absorption-dominated objects.  The rest of
the objects with no redshifts have low-\snr\ spectra such that no line
features could be measured, leaving their redshifts unknown.  For the
subset with redshifts, secure detection of lines, not necessarily
including $\fOII\lam3727$, were made that redshifts could be measured.
Not surprisingly, all the spectra in this subset had good enough \snr\
for detecting the \fOII\ equivalent width at the predefined
significance level; we used the criterion as mentioned above for
distinguishing low-\snr\ spectroscopic object from true non-detection.
The cluster membership was simply assumed for an object whose redshift
moved the central wavelength of its $\fOII\lam3727$ emission line into
the on-band filter passband.

\tab{tab.OIITrueDetectionsByGIClr} shows that the true detection rate
strongly depends on the \gi\ color of objects, generally confirming
the conclusions drawn by \citetalias{mlf00}.  Most $\gi < 1$ objects
turned out to be true \fOII\ emitters with very high successful
detection rate ($\simeq 85\%$).  One of the five false detections of
$\gi < 1$ objects was $\MgII\lam2798$ emission line of a $z\simeq0.87$
object; others were detection of $4000~\ang$ break from objects at
$z\simeq0.47$ ($1$) and at $z\simeq0.31$ ($3$).

For $1 < \gi < 2$ objects, the true detection rate was lower, in part
due to the stricter definition of true \fOII\ detection employed in
this analysis, requiring $W(\fOII) < -11~\ang$.  Five of the seven
cluster members classified into no \fOII\ emission category have
visually detectable \fOII\ emission, and the remaining two exhibited a
typical post-starburst spectrum with little or no \fOII\ emission; in
fact, strong Balmer absorption lines and a prominent $4000~\ang$ break
appeared common among this category.  The false detections of
non-member $1<\gi<2$ objects were caused by $4000~\ang$ break from the
objects at $z\simeq0.35$ ($3$) or $z\simeq0.39$ ($2$) and by unknown
reasons for the objects at $z\simeq0.62$ and $0.53$.

Most of $\gi > 2$ objects were false detections of early-type
galaxy spectra or possibly a few late-type K- or M-type stars
which showed few or no emission lines, and therefore their redshifts
were not measured.  In three $\gi > 2$ objects we detected \fOII\
emission line: one luminous object with a typical early-type galaxy
spectrum but also with emission lines---an elliptical galaxy hosting a
low-luminosity AGN; one very faint object with a spectrum in which
only \fOII, \fOIII, and Balmer lines could be visually identified; and
one object with a spectrum which appeared a superposition of two
different objects.

The results have shown that a narrowband technique is a very effective
method of selecting emission-line galaxies, but the true detection of
emission-line galaxies requires some constraints to be put on their
photometric properties.

\subsection{\fOII\ Line Flux Measurements}
\label{sec.OIILineFluxMeasurements}

\begin{figure}
\plotone{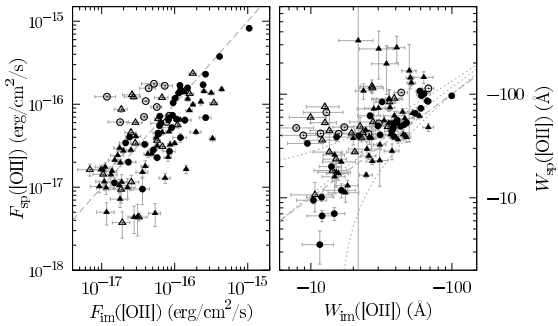}
\caption{ Comparison of $\fOII\lam3727$ emission line flux
(\emph{left}) and equivalent width (\emph{right}) measured in
\citetalias{mlf00} narrowband imaging and this spectroscopic
follow-up.  Circles and triangles denote LRIS and DEIMOS spectra,
respectively.  Filled and open data points denote on-band and on-edge
samples, respectively (\sect{sec.OIILineFluxMeasurements}).  The gray
dashed lines indicate the loci of equality.  For the right figure, the
dotted curves indicate a linear fit and $\pm 1\sigma$ bounds to the
on-band sample.  The final $\sigma$ was computed after one iteration
of $\pm 1\sigma$ clipping.  Upon fitting each data point was given a
weight of $1/\sigma^2$, where $\sigma$ is the sum in quadrature of the
two measurement errors in imaging and spectroscopic equivalent
widths. }
\label{fig.O2SpecVsImg}
\end{figure}

With the systematic errors in flux calibration
(\sect{sec.ObservationsAndDataReductions}), analysis using absolute
fluxes may be unreliable.  In \fig{fig.O2SpecVsImg}, $\fOII\lam3727$
flux and equivalent width measured from spectra and \citetalias{mlf00}
narrowband imaging are compared.  The comparisons of the two sets of
independent measurements are complicated by several factors.  First,
an on-band flux excess from narrowband imaging was only a lower limit
of a total integrated emission flux if at least a portion of the
redshifted \fOII\ line profile fell on the steeply declining portion
of the on-band filter transmission curve.  Second, a spectroscopic
flux of our multislit spectroscopic sample were only a lower limit to
the total \fOII\ emission flux from a galaxy due to an aperture
effect.  In this case, both the relative size of the aperture and the
galaxy and misalignment of an aperture with respect to the galaxy
center could produce noticeable systematics, because the composition
of the stellar populations covered by the aperture (e.g., bulge and
disk) may change significantly.  Since each $1''$-width slit was
supposed to be aligned to the center of each galaxy with a typical
angular size of $\sim 3''$--$4''$ in diameter, each spectrum was
spatially sampling only around a nuclear region extending over roughly
$\sim 10~\kpc$.  Third, in \citetalias{mlf00} only the continuum band
placed blueward of \fOII\ emission was used to estimate the continuum
flux density near \fOII\ emission.  This may cause discrepancy in the
\fOII\ flux and equivalent width measurements, as the \fOII\ emission
appears near the $4000~\ang$ break.

To see the effect of incomplete integration over a line profile in
narrowband photometry, we divided the sample into three categories.
Using the central wavelength $\lambda_c$ of redshifted \fOII\ emission
line profile, there are (1)~an \emph{on-band sample} consisting of
those objects with $5222.5~\ang \le \lambda_c < 5264.9~\ang$ for which
the transmission efficiency is over $0.7$; (2)~an \emph{on-edge
sample} either with $5197.0~\ang \le \lambda_c < 5222.5~\ang$ or with
$5264.9~\ang \le \lambda_c < 5292.0~\ang$; and (3)~an \emph{off-band
sample} either with $\lambda_c < 5197.0~\ang$ or with $\lambda_c \ge
5292.0~\ang$, where the transmission efficiency drops below $10^{-3}$.
Since the off-band sample are not cluster members under the definition
used in this paper, they are excluded from this analysis.

In \fig{fig.O2SpecVsImg} the major outliers above the line of equality
tend to be of on-edge sample, and the narrowband \fOII\ measurements
generally are smaller than those of spectra.  This discrepancy can be
easily attributed to the incomplete integration over a \fOII\ line
profile, i.e., the narrowband flux gets underestimated while the
off-band continuum remains relatively unchanged, giving smaller
equivalent width.  Within the on-band sample for which the narrowband
flux is a good estimate for the total \fOII\ emission flux, the
spectroscopic flux generally underestimates the \fOII\ flux due to
aperture effect.

\begin{figure}
\plotone{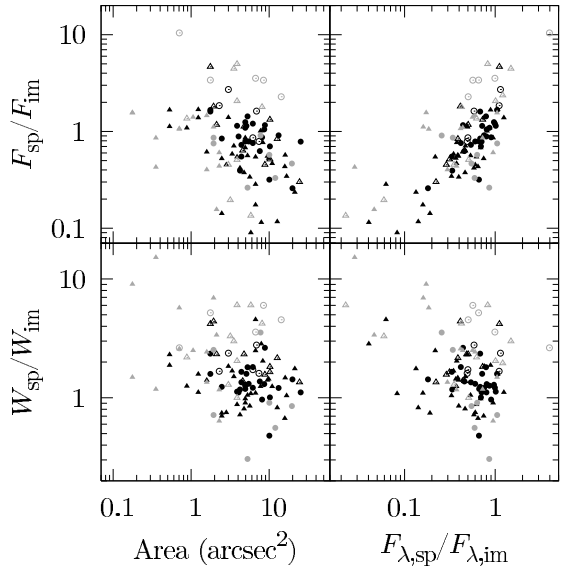}
\caption{ Distribution of discrepancies in imaging and spectroscopic
measurement of \fOII\ flux (\emph{top row}) and equivalent width
(\emph{bottom row}) by the source isophotal angular size (\emph{left
column}) and the discrepancy in the \fOII\ continuum flux densities
(\emph{right column}) determined by \citetalias{mlf00} narrowband
imaging and this spectroscopic studies.  The measurements for which
the $\snr < 5$ for on-band imaging flux excess and spectroscopic flux
density at the line center are grayed out.  The definition of symbols
are identical as in \fig{fig.O2SpecVsImg}. }
\label{fig.FWDiscrepancy}
\end{figure}

To see if the relative size of the aperture and galaxy, misalignment,
or the discrepancy in the determination of narrowband continuum flux
density near \fOII\ contributes significantly to systematics, we
plotted the ratio of spectroscopic to \citetalias{mlf00} narrowband
imaging measurements against the source isophotal area from the
\citetalias{mlf00} narrowband survey and the ratio of spectroscopic to
imaging \fOII\ continuum flux densities in \fig{fig.FWDiscrepancy}.

If the relative size between the aperture and galaxy was a major cause
of discrepancy between narrowband and spectroscopic measurements, we
would expect the agreement between the two measurements to be better
for smaller galaxies.  The upper left plot of \fig{fig.FWDiscrepancy}
suggests the flux loss for larger galaxies is indeed significant.  The
upper right plot shows a strong positive correlation between the
discrepancy in \fOII\ flux and the continuum flux density
measurements.  Notably our flux-calibrated spectra could underestimate
\emph{both} \fOII\ emission flux and continuum flux density by up to a
factor of $\sim 3$, ignoring possible outliers for which the ratios
were $\la 0.3$.  The linear correlation, which appears to be followed,
is expected if the discrepancy between narrowband and spectroscopic
measurements is dominated by the systematic error in flux calibration.
The upper plots appear to suggest the presence of aperture effect and
systematic error in flux calibration.

The corresponding trends are harder to observe in lower plots,
comparing spectroscopic and narrowband equivalent widths.
Spectroscopic \fOII\ equivalent widths are slightly greater in general
than those from narrowband measurements.  The sloping continuum of the
Balmer break around $\fOII\lam3727$ usually causes the continuum
blueward of the line to have lower flux density than that of the
redward continuum.  We suspected this underestimate of continuum flux
density at \fOII\ might cause narrowband \fOII\ emission equivalent
widths to become systematically stronger, but the effect appears
negligible.

If we suppose the underestimate of spectroscopic flux is caused by the
aperture effect and/or misalignment, it is only to the extent that the
underlying stellar population covered by the aperture does not change
substantially because the equivalent width measurements show little
trend with the source isophotal area and the discrepancy of their
continuum flux densities.  However, the plots do indicate a few
low-\snr\ measurements yield very large equivalent widths, probably
due to poorly defined continuum flux density.

We conclude that our flux-calibrated spectra can yield underestimates
by a factor of up to $\sim 3$ for absolute physical quantities due to
the combination of aperture effect, misalignment, and systematic error
in flux-calibration but only to the extent that the equivalent widths
remain robust.  We also note there also was a slight seeing variation
which would affect flux measurements.  At least some of unusually
large spectroscopic \fOII\ equivalent widths are likely to be caused
by large uncertainties in the continuum flux density near \fOII\
emission line.

\section{Properties of the Cluster \fOII\ Emitters}
\label{sec.PropertiesOfTheClusterOIIEmitters}

\subsection{AGN Diagnostics}
\label{sec.AGNDiagnostics}

\begin{figure}
\plotone{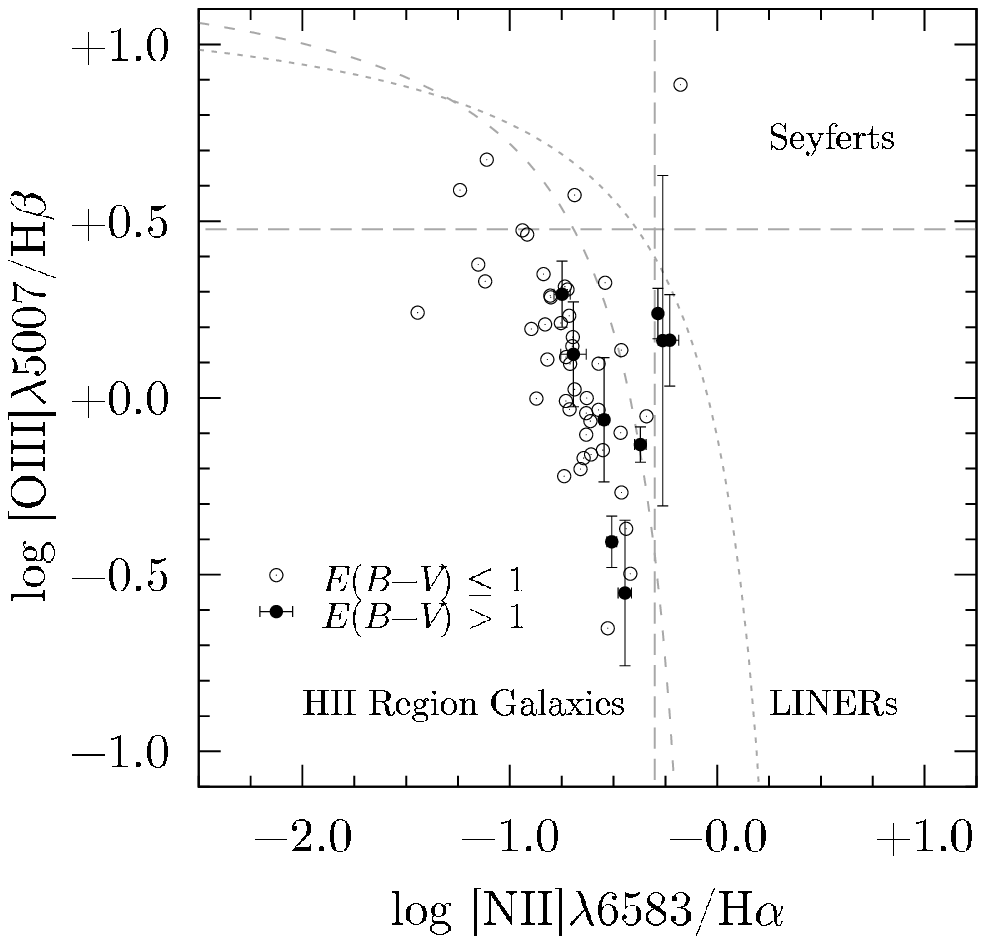}
\caption{ Line ratio diagnostic diagram for $\fNII\lam6584$
vs.~$\fOIII\lam5007$ normalized to the nearest Balmer emission line.
The classifications indicated follow the scheme of \citet{ho96}, but
the cut was made at $\fNII/\Ha = 0.5$ instead of $0.6$.  The
demarcation defined by \citet{kau03} (dashed curve) and \citet{kew01}
(dotted curve) are also shown for comparison.  Galaxies with high dust
reddening [$\EBV > 1$] are separately indicated with their error bars.
}
\label{fig.LogO3HbVsLogN2Ha}
\end{figure}

Some of the observed \fOII\ emission could be from active galactic
nuclei (AGNs).  In \fig{fig.LogO3HbVsLogN2Ha} we show a line ratio
diagnostics diagram for AGN activity using the scheme outlined in
\citet{ho96}.  Only the objects with measured line fluxes necessary to
compute the diagnostic ratios are in the diagram.  A similar diagram
using $\fSII\lam\lam6716,31$ doublet was not made since the lines were
outside the observed window or severely compromised by sky lines.  We
deviated from their scheme in that the distinction between AGN-like
and \HII\ region-like galaxies was made at $\fNII\lam6583 / \Ha =
0.5$, instead of $0.6$, to include objects whose spectra apparently
have broad emission lines upon visual inspection.  This criterion
identifies a total of $7$ galaxies to be AGN-like, $3$ of which cannot
be plotted in \fig{fig.LogO3HbVsLogN2Ha} for their lack of either
\fOIII\ or \Hb\ measurements.  For comparison, more recent demarcation
from \citet{kau03} and \citet{kew01} are also shown.  We see that the
AGN identification is indeed sensitive to the choice of demarcation,
yet we keep the above criterion since it does remove objects with
visually identifiable broad emission lines.  The diagnostic line
fluxes are not corrected for reddening, but taking a ratio to the
nearest Balmer line, e.g., $\fOIII\lam5007 /\Hb$ and $\fNII\lam6583
/\Ha$ minimizes the effect of reddening; applying reddening correction
did not change the results significantly.  In order to include more
objects in this diagnosis, however, we deviated from our convention of
using the best \snr\ spectrum and used a secondary spectrum if AGN
diagnostics cannot be done otherwise
(\sect{sec.ObservationsAndDataReductions}); if a spectrum from a
repeat observation at a lower \snr\ but with both $\fNII\lam6583$ and
\Ha\ measurements exist, we used that spectrum to do the AGN
diagnostic.

As far as their \gi\ colors are concerned, it is hard to distinguish
AGN-like galaxies from the rest of the \fOII\ emitters, although they
are found mostly toward the bright end in the color-magnitude diagram.
It should also be noted that our AGN-like galaxies are not detected in
previous X-ray observations; we will discuss the implication of these
results in a forth-coming paper in the series.  Most cluster \fOII\
emitters are classified as normal star-forming, \HII\ region-like
galaxies.  Where appropriate, $7$ objects identified to be AGN-like
will be removed from subsequent analysis of star-forming galaxies.

\subsection{Derivation of \EBV}

We derived reddening \EBV\ from the Balmer decrement of either
$\Ha/\Hb$ or $\Hc/\Hb$ emission line flux ratio, depending on the
availability of the lines, with the presumed theoretical line ratios
of $(\Ha/\Hb)_0 = 2.85$ and $(\Hc/\Hb)_0 = 0.469$ for Case B
recombination at $T_{\rm e} = 10^4~{\rm K}$ and $n_{\rm e} =
10^4~\cm^{-3}$ from \citet{ost89}.  The relation to convert the
observed flux $F$ into the intrinsic flux $F_0$ is
\[
F_0 = F  10^{0.4 E(B-V) k^e(\lambda)} \ ,
\]
where $k^e(\lam)$ is the extinction curve described in \citet{cal00}.
Being derived from a ratio of nebular emission lines, \EBV\ therefore
is the amount of reddening in the nebular component, which is
generally greater by a factor of $\sim 2$ than that measured in the
stellar component \citep[e.g.,][]{cal94}.

Unfortunately, the spectral region around the redshifted \Hb\ often
suffer from a telluric absorption feature near $\sim 6870~\ang$,
so the quantities derived partially from \Hb\ measurements tend to
have large uncertainties.  We emphasize that the high uncertainty in
our \Hb\ measurements is largely the consequence of deriving the
uncertainty in line flux from the RMS of Gaussian fit, but it is the
\Hb\ \emph{absorption profile} which gets partially contaminated by
the telluric feature at the redshift of interest.  The Gaussian
fitting itself was done with weighting by inverse variance, which
includes the uncertainty due to sky subtraction
(\sect{sec.SpectralLineMeasurements}).  Therefore we may have
overestimated the uncertainty in emission flux, which is almost never
compromised by the same sky line.  We also make clear that at \Hb\ the
emission and absorption profiles could almost always be fit
separately; therefore the derived \EBV\ do not get overestimated from
possible underestimation of underlying absorption at \Hb\
(\sect{sec.GaussianFittingMethod}).

\subsection{$\fOII/\Ha$ Ratio \& Star Formation Rate}
\label{sec.OIIHaRatioAndSFR}

\begin{figure*}
\plotone{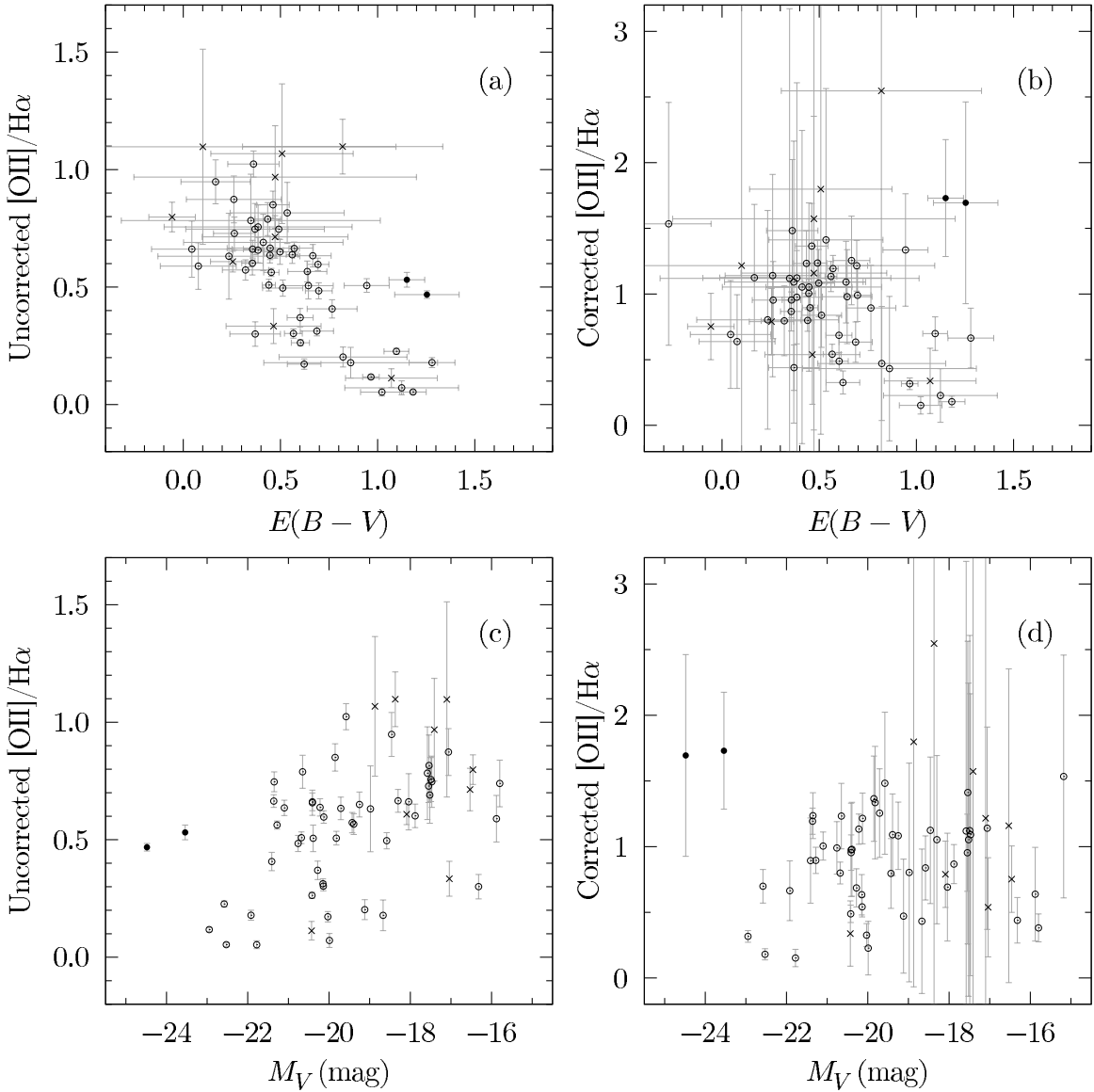}
\caption{ The distribution of emission line flux ratio $\fOII/\Ha$ as
a function of reddening \EBV\ derived from a Balmer decrement,
(\emph{a}) before and (\emph{b}) after extinction correction.  The
ratio $\fOII/\Ha$ as a function of absolute $V$ magnitude (\emph{c})
before and (\emph{d}) after extinction correction are also plotted.
Open and filled circles are $\fOII/\Ha$ of normal, \HII\ region-like
galaxies and AGN-like objects, respectively
(\sect{sec.AGNDiagnostics}).  Crosses are the objects whose
classification by the line ratio diagnostics could not be carried out.
}
\label{fig.O2HaVsEBV}
\end{figure*}

The empirical correlation between $\fOII\lam3727$ emission line and
\Ha\ fluxes calibrated by \citet{gal89} and \citet{ken92} is often
used to estimate star formation rate (SFR) of distant galaxies, since
the \fOII\ doublet is the strongest emission line in the blue that
remains available within the optical range for high-$z$ galaxies.  In
\fig{fig.O2HaVsEBV} we show the dependence of the observed $\fOII/\Ha$
line flux ratio on the amount of reddening \EBV\ and galaxy
luminosity.  The absolute $V$-band magnitudes, $M_V$ (on the Vega
scale), are derived from the reddening-corrected $g$ and $i$
photometry using software package kcorrect
\citep[v4\_1\_4;][]{bla03}.\footnote{The objects for which we could
not derive reddening \EBV\ from their line ratio are therefore removed
from the current analysis.}  The relative independence of $\fOII/\Ha$
on dust reddening after correction is one of the basic assumptions for
the reliability of $\fOII\lam3727$ emission as a star formation
indicator.  To compare with previous studies, we compute the mean
$\fOII/\Ha$ for the \fOII\ emitters identified to be regular \HII\
region-like galaxies (\sect{sec.AGNDiagnostics}).  Before reddening
correction, the mean ratio of $\fOII/\Ha = 0.56 \pm 0.01$ is
comparable to $0.57 \pm 0.06$ and $0.73 \pm 0.03$ measured by
\citet{ken92} and \citet{jan01}, respectively.  After correction, the
ratio becomes $0.88 \pm 0.08$ for our sample, which is slightly lower
than $1.1 \pm 0.1$ and $1.2 \pm 0.3$ obtained by the aforementioned
studies.  The ranges of luminosities in these samples are roughly
consistent with each other.

The range of \EBV\ in \fig{fig.O2HaVsEBV} implies a significant
fraction of \fOII\ emitters suffers high degree of dust reddening.
Whereas the galaxies \citet{kew04} used for their calibration of SFR
mostly have the extinction of $\EBV \la 0.5$, a significant fraction
of our cluster \fOII\ emitting sample has $\EBV > 0.5$.  It is well
known that optical diagnostic methods such as $\fOII\lam3727$ emission
line flux severely underestimate the amount of star-formation
activities in presence of high dust reddening, when no extinction
correction is made.  Although a very strong trend with \EBV\ is
removed by reddening correction, the ratio $\fOII/\Ha$ for very dusty
galaxies remains about a factor of two lower than the mean for
the entire sample.  Indeed the low $\fOII/\Ha$ ratio of the
high-\EBV\ galaxies caused our mean $\fOII/\Ha$ to be lower than the
ratios measured by previous studies mentioned above.  Based on the
figure, the reliability of the widely used SFR calibration between
\fOII\ and \Ha\ emission lines \citep[e.g.,][]{ken98} may not extend
to very dusty galaxies in general, e.g., $\EBV \ga 0.7$, which appear
abundant in high-$z$ universe and have been reported to show smaller
$\fOII/\Ha$ as well \citep{hic02}.  Furthermore, the uncorrected
$\fOII/\Ha$ shows a modest trend with luminosity in
\fig{fig.O2HaVsEBV}.  Although reddening correction appears to remove
the luminosity trend, a few very low-$\fOII/\Ha$ galaxies do not move
closer to the mean ratio.  In fact, these are $\EBV > 0.5$ galaxies,
and their distribution in luminosity does not appear to show a strong
trend.

As reported by \citet{jan01}, the relations among the $\fOII/\Ha$
ratio, dust reddening, and luminosity appear coupled, yet our cluster
\fOII\ emitting sample in \fig{fig.O2HaVsEBV} shows that $\fOII/\Ha$
has the strongest trend with the amount of dust reddening, and for
very dusty galaxies, their $\fOII/\Ha$ ratio remains lower than the
locally measured means regardless of their luminosity.  The primary
reason for this trend may be the dependence of $\fOII/\Ha$ ratio on
the ionization state and the gas phase metallicity of emission nebula
as reported in metal-rich \citep{kew04} and metal-poor galaxies
\citep{mou05}.  To study the relevance of $\fOII/\Ha$ ratio further,
it is necessary to constrain and consider the gas-phase metallicity
and ionization state of galaxies in our sample.

\subsection{\fOII\ Emission \& Luminosity}

The mass associated with the stellar component of a galaxy is better
traced in redder passbands, which measure flux from the part of galaxy
spectrum relatively insensitive to the recent star formation history
and also suffers less dust extinction.  The reddest imaging in our
survey is Sloan $i$, so we could take the observed $i$ magnitude as
the quantitative estimate for the underlying stellar mass, or more
generally the size, of a galaxy.  In the cluster rest-frame the Sloan
$i$ filter really samples the $V$-band light of galaxy spectrum.  Thus
$i$ magnitude is not the best proxy for the underlying stellar
population that dominates the stellar mass composition of a galaxy.
An instantaneous burst model of \citet{bru03} shows that at $10~\Myr$
a starburst fraction of only $\sim 1\%$ can make a dominant
contribution to the visual spectrum at $\sim 5500~\ang$ on top of
$10~\Gyr$-old stellar population.  Although we use
extinction-corrected absolute $V$ magnitude to estimate the intrinsic
luminosity or size of galaxies, some caution should be exercised to
interpret our results, as the quantities are highly dependent on their
SFHs.

\begin{figure}
\plotone{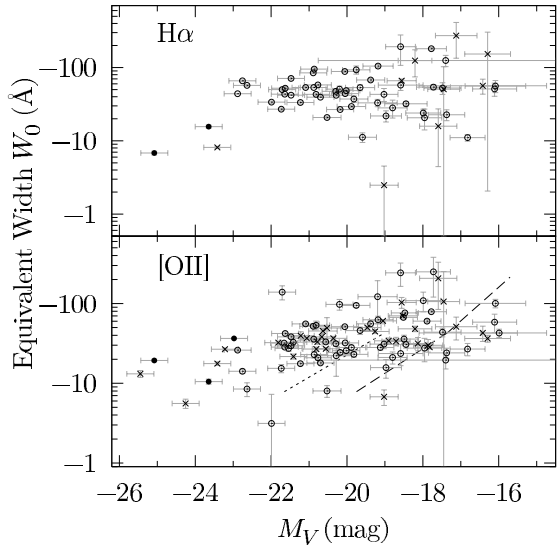}
\caption{ Rest-frame \fOII\ (\emph{bottom}) and \Ha\ (\emph{top})
equivalent width of the cluster \fOII\ emitters as a function of their
absolute $V$ magnitude.  Symbols are as in \fig{fig.O2HaVsEBV}.
Dashed and dotted lines in the bottom plot roughly indicate the
sensitivity limits of \citetalias{mlf00} narrowband survey for
galaxies extending over 3\farcs8 (9 pixels) and 1\farcm4 (200 pixels),
respectively.  For some galaxies \Ha\ was outside the observed
spectral range while their \fOII\ equivalent width could be measured.}
\label{fig.W0OIIHaVsImag}
\end{figure}

Previous studies have noted that the \fOII\ equivalent widths tend to
anti-correlate with luminosity \citep{cow96,hog98,bla00,cha02}.  In
\fig{fig.W0OIIHaVsImag}, we observe similar trend in \fOII\ and \Ha\
equivalent widths (measured with the flux-summing method).  Since the
completeness limit of \citetalias{mlf00} survey was $W(\fOII) \simeq
11~\ang$ at $i \simeq 23$ ($M_V \simeq -19$), the lack of
intrinsically faint weak \fOII\ emitters is partly a result of
selection effect.  A few faint galaxies have ill-defined, low-\snr\
continuum which could produce unusually large \fOII\ equivalent widths
(\fig{fig.FWDiscrepancy}), but our measurement and associated
uncertainty of equivalent widths via the flux-summing method
(\sect{sec.SpectralLineMeasurements}) generally is robust even with
low-\snr\ spectra, i.e., it does not show a systematic bias toward
larger equivalent widths unlike when a Gaussian is fitted to a
low-\snr\ line profile.  We checked the robustness by artificially
degrading a high-\snr\ spectrum, with which a robust measurement of
equivalent widths can be made, and actually measuring equivalent
widths by the flux-summing method.

Although very strong \fOII\ equivalent widths, e.g., $W(\fOII) \la
-100~\ang$, are generally observed with large uncertainties in faint
galaxies below the survey completeness limit, the anti-correlation
between the emission equivalent width and galaxy luminosity is clearly
observed in \fOII\ emission but appears almost non-existent in \Ha\
emission.  At a glance, there is a strong suppression of emission
equivalent widths in very luminous galaxies.  In fact, some of the
most luminous galaxies, i.e., $M_V < -22$, are classified as AGN-like,
whereas few very luminous galaxies with high \fOII\ emission
equivalent width exist in the same luminosity range.  That AGNs are
preferentially hosted in very luminous objects is not surprising, but
it is not clear that the absence of low-luminosity AGNs (i.e., $M_V >
-22$) is real or is due to selection bias.  It is known that the
strength of \fOII\ and \Ha\ emission, when excited by AGNs, correlates
tightly with the AGN luminosity.  Therefore if the observed mean
equivalent width of the luminous AGN-like galaxies, of order of $\sim
10~\ang$, were to set roughly the maximum equivalent width, most AGNs
of lower luminosity would be expected to drop out of our detection
limit.

A potentially intriguing interpretation of \fig{fig.W0OIIHaVsImag} is
a manifestation of so-called \emph{downsizing effect}, which refers to
observational evidence for the redshift evolution of characteristic
galaxy luminosity above which star formation activities are suppressed
\citep{cow96}.  Such characteristic luminosity at which the
suppression of vigorously star-forming galaxies evolves toward less
luminous end at lower redshifts.  Now, AGN activity has been shown by
simulations to be more effective in suppressing star formation in
major galaxy mergers than those involving no nuclear activities
\citep{spr05}.  Visual inspection of our images indicates the AGN-like
galaxies to be located in among the densest regions in terms of
projected surface number density of galaxies.  Our imaging does not
have the resolution to clearly see the signature of mergers, yet the
distribution of AGN-like galaxies in our sample might suggest the
implicit role of AGN activities in regulating star formation activity
in large galaxies.  While the interpretation is highly speculative and
there have been little direct observational support, recently a number
of studies which provide circumstantial evidence in support for this
scenario have been emerging.  For example, the buildup of cluster red
sequence appears to start from the high-luminosity end \citep{del04},
while the number of post-starburst galaxies appear to host AGN
activities (C. Tremonti et al. 2006, private communication).  In view
of these recent findings, the weaker \fOII\ equivalent widths found in
high-luminosity AGN-like galaxies in our sample are at least
consistent with AGNs suppressing cluster star-forming activities.  We
will revisit this argument by analyzing the star formation histories,
local environments, and archival \HST\ morphology of the cluster
\fOII\ emitters in the forthcoming paper in the series.

Putting the high-luminosity \fOII\ emitters aside, interpreting such a
luminosity trend as in \fig{fig.W0OIIHaVsImag} is not trivial, since
both metallicity and the amount of dust extinction generally show a
positive trend with galaxy luminosity \citep[e.g.,][]{hec98,tre04} and
affect star formation diagnostics like emission equivalent width.  In
particular, \citet{cha02} found in their sample of local galaxies
drawn from the Stromlo-APM redshift survey that the star formation
rate per unit luminosity showed little dependence on luminosity, when
dust extinction was properly accounted for.  We see an indication of
significance of dust extinction when \fOII\ and \Ha\ emission
equivalent widths are compared, i.e., \fOII\ equivalent widths show a
stronger trend with luminosity than \Ha\ equivalent widths.

\begin{figure}
\plotone{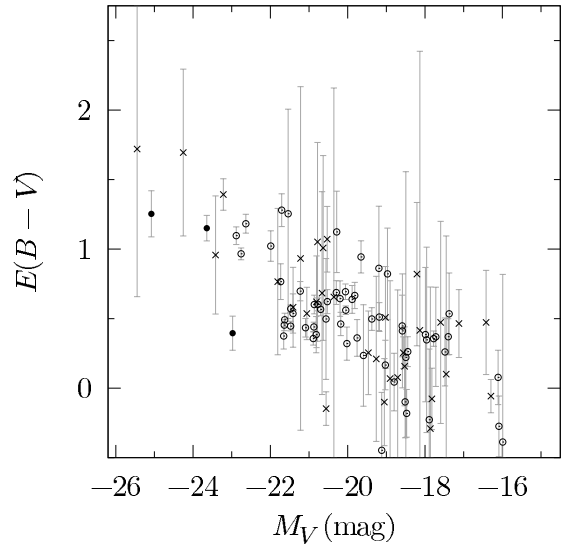}
\caption{ Reddening \EBV\ of cluster \fOII\ emitters as a function of
their absolute $V$ magnitude.  The symbols have the same meaning as in
\fig{fig.O2HaVsEBV}. }
\label{fig.EBVVsIMag}
\end{figure}

The equivalent width of nebular emission line such as \fOII\ and \Ha\
is essentially a ratio of the total emission flux from an ensemble of
localized \HII\ regions to the continuum emission from stellar
population more widely distributed throughout a galaxy.  It is
empirically known that nebular emission lines suffer more dust
extinction than the stellar continuum \citep{cal94}.  Therefore dust
extinction generally reduces the strength of emission equivalent
width, and that effect is selectively stronger at shorter wavelengths,
i.e., blue emission line suffers more from this effect due to the
wavelength dependence of dust absorption.  In \fig{fig.EBVVsIMag} we
see that the reddening \EBV\ of cluster \fOII\ emitters shows a trend
with their luminosity, especially at the most luminous end; the
\citeauthor{jan01} sample showed a similar trend, although their
galaxies were not as reddened as ours.  Thus the observed trend of
decreasing emission equivalent widths with galaxy luminosity in
\fig{fig.W0OIIHaVsImag} could be enhanced by the greater amount of
dust extinction in high luminosity galaxies.  Furthermore, the
positive luminosity trend of \EBV\ points to the importance of
metallicity in decoupling the $\fOII/\Ha$ ratio trend with luminosity
and reddening (\fig{fig.O2HaVsEBV}), since the dust abundance is
expected to scale with metallicity, and the $\fOII/\Ha$ ratio of
nearby galaxies has been shown to correlate nontrivially with the
metallicity as well as ionization state of emission nebula
\citep{kew04,mou05}.  With rather large uncertainties, we may not
convincingly say if the low-luminosity galaxies in our sample are more
reddened than the local counterparts.  Yet a preliminary analysis of the \Hd-strong galaxy fraction in our sample shows a higher
incidence of dusty starburst galaxies in this cluster  compared to
a local field sample, probably due to galaxy-galaxy interactions.  In
a future paper we also wish to explore the metallicity effect on
$\fOII/\Ha$.

\section{Summary}
\label{sec.Summary}

We have reported on a spectroscopic follow-up confirmation of faint
\fOII\ emitters in \object{Abell 851} catalogued by
\citetalias{mlf00}.  The Keck~I/LRIS and Keck~II/DEIMOS spectra have
confirmed $\fOII\lam3727$ emission lines of the narrowband-selected
cluster \fOII\ emitters at a $\approx 85\%$ success rate for faint ($i
\la 25$) blue ($\gi < 1$) galaxies.  The rate for the successful
detection of \fOII\ emission is a strong function of galaxy color,
generally proving the efficacy of narrowband technique supplemented
with broadband colors in wide-field search for faint emission-line
objects.  Balmer decrement-derived reddening measurements show these
galaxies are very dusty, with a significant fraction of the population
showing $\EBV \ga 0.5$.  Even after correcting for dust extinction,
the $\fOII/\Ha$ ratio for very high \EBV\ galaxies remains generally
lower by a factor of $\sim 2$ than the mean $\fOII/\Ha$ ratios
reported by the studies of nearby galaxies.  A slight luminosity trend
of $\fOII/\Ha$ can be mostly removed by reddening correction, yet the
emission line flux ratio for high-\EBV\ galaxies remains much lower
regardless of their luminosity.  In a future paper, we will further
test the metallicity dependence of $\fOII/\Ha$ ratio, which the
previous studies of nearby galaxies have shown to be strong.  The
luminosity trend of \fOII\ and \Ha\ emission equivalent width
strength, coupled with the abundance of luminous AGN-like galaxies, is
consistent with AGNs suppressing star-forming activities for massive
cluster galaxies, although further studies specifically addressing
this issue need to be carried out.

\acknowledgments T.S. thanks Michael Cooper and Jeff Newman for
assistance in the use of DEEP2 DEIMOS Data pipeline.  T.S. also thanks
Marcin Sawicki, Tommaso Treu, and Jong-Hak Woo for useful
conversations.  We thank an anonymous referee for valuable comments
which improved the presentation of this paper.
Financial support was provided by the David and Lucille Packard
Foundation and the Alfred P. Sloan Foundation.  This research has made
use of the NASA/IPAC Extragalactic Database (NED), which is operated
by the Jet Propulsion Laboratory, California Institute of Technology,
under contract with the National Aeronautics and Space Administration.
This research has made use of the NASA Astrophysics Data System
abstract service.
The authors wish to recognize and acknowledge the very significant
cultural role and reverence that the summit of Mauna Kea has always
had within the indigenous Hawaiian community.  We are most fortunate
to have the opportunity to conduct observations from this mountain.

{\it Facilities:} \facility{Keck:I (LRIS)}, \facility{Keck:II (DEIMOS)}


\begin{thebibliography}{}

\bibitem[Abraham et al.(1996)]{abr96b} Abraham, R.~G., van den
Bergh, S., Glazebrook, K., Ellis, R.~S., Santiago, B.~X., Surma, P., \&
Griffiths, R.~E.\ 1996, \apjs, 107, 1
\bibitem[Balogh et al.(1997)]{bal97} Balogh, M.~L., Morris,
S.~L., Yee, H.~K.~C., Carlberg, R.~G., \& Ellingson, E.\ 1997, \apjl, 488,
L75
\bibitem[Blanton \& Lin(2000)]{bla00} Blanton, M., \& Lin,
H.\ 2000, \apjl, 543, L125
\bibitem[Blanton et al.(2003)]{bla03} Blanton, M.~R., et al.\
2003, \aj, 125, 2348
\bibitem[Bruzual \& Charlot(2003)]{bru03} Bruzual, G., \&
Charlot, S.\ 2003, \mnras, 344, 1000
\bibitem[Butcher \& Oemler(1984)]{but84} Butcher, H., \&
Oemler, A.\ 1984, \apj, 285, 426
\bibitem[Calzetti et al.(1994)]{cal94} Calzetti, D., Kinney,
A.~L., \& Storchi-Bergmann, T.\ 1994, \apj, 429, 582
\bibitem[Calzetti et al.(2000)]{cal00} Calzetti, D., Armus,
L., Bohlin, R.~C., Kinney, A.~L., Koornneef, J., \& Storchi-Bergmann, T.\
2000, \apj, 533, 682
\bibitem[Charlot et al.(2002)]{cha02} Charlot, S., Kauffmann,
G., Longhetti, M., Tresse, L., White, S.~D.~M., Maddox, S.~J., \& Fall,
S.~M.\ 2002, \mnras, 330, 876
\bibitem[Cowie et al.(1996)]{cow96} Cowie, L.~L., Songaila,
A., Hu, E.~M., \& Cohen, J.~G.\ 1996, \aj, 112, 839
\bibitem[De Filippis et al.(2003)]{def03} De Filippis, E.,
Schindler, S., \& Castillo-Morales, A.\ 2003, \aap, 404, 63
\bibitem[De Lucia et al.(2004)]{del04} De Lucia, G., et al.\
2004, \apjl, 610, L77
\bibitem[Dressler(1980)]{dre80} Dressler, A.\ 1980, \apj,
236, 351
\bibitem[Dressler \& Gunn(1992)]{dre92} Dressler, A., \&
Gunn, J.~E.\ 1992, \apjs, 78, 1
\bibitem[Dressler et al.(1999)]{dre99} Dressler, A., Smail,
I., Poggianti, B.~M., Butcher, H., Couch, W.~J., Ellis, R.~S., \& Oemler,
A.~J.\ 1999, \apjs, 122, 51
\bibitem[Faber et al.(2003)]{fab03} Faber, S.~M., et al.\
2003, \procspie, 4841, 1657
\bibitem[Gallagher, Hunter, \& Bushouse(1989)]{gal89}
Gallagher, J.~S., Hunter, D.~A., \& Bushouse, H.\ 1989, \aj, 97, 700
\bibitem[Heckman et al.(1998)]{hec98} Heckman, T.~M., Robert,
C., Leitherer, C., Garnett, D.~R., \& van der Rydt, F.\ 1998, \apj, 503,
646
\bibitem[Hicks et al.(2002)]{hic02} Hicks, E.~K.~S., Malkan,
M.~A., Teplitz, H.~I., McCarthy, P.~J., \& Yan, L.\ 2002, \apj, 581, 205
\bibitem[Ho(1996)]{ho96} Ho, L.~C.\ 1996, ASP
Conf.~Ser.~103: The Physics of Liners in View of Recent Observations, 103
\bibitem[Hogg et al.(1998)]{hog98} Hogg, D.~W., Cohen, J.~G.,
Blandford, R., \& Pahre, M.~A.\ 1998, \apj, 504, 622
\bibitem[Jansen et al.(2001)]{jan01} Jansen, R.~A., Franx,
M., \& Fabricant, D.\ 2001, \apj, 551, 825
\bibitem[Kauffmann et al.(2003)]{kau03} Kauffmann, G., et
al.\ 2003, \mnras, 346, 1055
\bibitem[Kauffmann(1995a)]{kau95a} Kauffmann, G.\ 1995a, \mnras,
274, 153
\bibitem[Kauffmann(1995b)]{kau95b} Kauffmann, G.\ 1995b, \mnras,
274, 161
\bibitem[Kennicutt(1992)]{ken92} Kennicutt, R.~C.\ 1992,
\apj, 388, 310
\bibitem[Kennicutt(1998)]{ken98} Kennicutt, R.~C.\ 1998,
\araa, 36, 189
\bibitem[Kewley et al.(2004)]{kew04} Kewley, L.~J., Geller,
M.~J., \& Jansen, R.~A.\ 2004, \aj, 127, 2002
\bibitem[Kewley et al.(2001)]{kew01} Kewley, L.~J., Dopita,
M.~A., Sutherland, R.~S., Heisler, C.~A., \& Trevena, J.\ 2001, \apj, 556,
121
\bibitem[Kodama et al.(2001)]{kod01} Kodama, T., Smail, I.,
Nakata, F., Okamura, S., \& Bower, R.~G.\ 2001, \apjl, 562, L9
\bibitem[Kodama et al.(2005)]{kod05} Kodama, T., et al.\
2005, \pasj, 57, 309
\bibitem[Le Floc'h et al.(2005)]{lef05} Le Floc'h, E., et
al.\ 2005, \apj, 632, 169
\bibitem[Lilly et al.(1996)]{lil96} Lilly, S.~J., Le Fevre,
O., Hammer, F., \& Crampton, D.\ 1996, \apjl, 460, L1
\bibitem[Lotz et al.(2003)]{lot03} Lotz, J.~M., Martin,
C.~L., \& Ferguson, H.~C.\ 2003, \apj, 596, 143
\bibitem[Madau et al.(1996)]{mad96} Madau, P., Ferguson,
H.~C., Dickinson, M.~E., Giavalisco, M., Steidel, C.~C., \& Fruchter, A.\
1996, \mnras, 283, 1388
\bibitem[Martin et al.(2000)]{mlf00} Martin, C.~L., Lotz, J.,
\& Ferguson, H.~C.\ 2000, \apj, 543, 97 \citepalias{mlf00}
\bibitem[Massey et al.(1988)]{mas88} Massey, P., Strobel, K.,
Barnes, J.~V., \& Anderson, E.\ 1988, \apj, 328, 315
\bibitem[McCall et al.(1985)]{mcc85} McCall, M.~L., Rybski,
P.~M., \& Shields, G.~A.\ 1985, \apjs, 57, 1
\bibitem[Moore et al.(1999)]{moo99} Moore, B., Lake, G.,
Quinn, T., \& Stadel, J.\ 1999, \mnras, 304, 465
\bibitem[Moran et al.(2005)]{mor05} Moran, S.~M., Ellis,
R.~S., Treu, T., Smail, I., Dressler, A., Coil, A.~L., \& Smith, G.~P.\
2005, \apj, 634, 977
\bibitem[Mouhcine et al.(2005)]{mou05} Mouhcine, M., Lewis,
I., Jones, B., Lamareille, F., Maddox, S.~J., \& Contini, T.\ 2005, \mnras,
362, 1143
\bibitem[Oke et al.(1995)]{oke95} Oke, J.~B., et al.\ 1995,
\pasp, 107, 375
\bibitem[Osterbrock(1989)]{ost89} Osterbrock, D.~E.\ 1989, Astrophysics
of Gaseous Nebulae and Active Galactic Nuclei
(Mill Valley: Univ. Science Books)
\bibitem[Schindler et al.(1998)]{schi98} Schindler, S.,
Belloni, P., Ikebe, Y., Hattori, M., Wambsganss, J., \& Tanaka, Y.\ 1998,
\aap, 338, 843
\bibitem[Schindler \& Wambsganss(1996)]{schi96} Schindler, S.,
\& Wambsganss, J.\ 1996, \aap, 313, 113
\bibitem[Schlegel et al.(1998)]{sch98} Schlegel, D.~J.,
Finkbeiner, D.~P., \& Davis, M.\ 1998, \apj, 500, 525
\bibitem[Springel et al.(2005)]{spr05} Springel, V., Di
Matteo, T., \& Hernquist, L.\ 2005, \mnras, 361, 776
\bibitem[Tasitsiomi et al.(2004)]{tas04} Tasitsiomi, A.,
Kravtsov, A.~V., Gottl{\" o}ber, S., \& Klypin, A.~A.\ 2004, \apj,
607, 125
\bibitem[Tremonti et al.(2004)]{tre04} Tremonti, C.~A., et
al.\ 2004, \apj, 613, 898
\bibitem[Treu et al.(2003)]{tre03} Treu, T., Ellis, R.~S.,
Kneib, J.~P., Dressler, A., Smail, I., Czoske, O., Oemler, A., \&
Natarajan, P.\ 2003, \apj, 591, 53
\bibitem[Zhao et al.(2003)]{zha03} Zhao, D.~H., Mo, H.~J.,
Jing, Y.~P., B{\" o}rner, G.\ 2003, \mnras, 339, 12

\end{thebibliography}
\end{document}